\documentclass[10pt,preprint]{aastex}
\begin{document}

\title{Cryogenic tests of volume-phase holographic gratings: results at
100 $K$}

\author{Naoyuki Tamura\altaffilmark{1}, Graham J. Murray, Peter Luke,
Colin Blackburn, David J. Robertson, Nigel A. Dipper, Ray M. Sharples,
\& Jeremy R. Allington-Smith}

\affil{Centre for Advanced Instrumentation, Department of Physics,
University of Durham, South Road, Durham, DH1 3LE, UK}

\altaffiltext{1}{Current address: Subaru telescope, 650 North A'ohoku
Place, Hilo, HI 96720, USA. E-mail: naoyuki@naoj.org}

\begin{abstract}
We present results from cryogenic tests of Volume-Phase Holographic
(VPH) gratings at $\sim$ 100 $K$. The aims of these tests are to see
whether the diffraction efficiency as a function of wavelength is
significantly different at a low temperature from that at room
temperature and to see how the performance of a VPH grating is affected
by a number of thermal cycles. We have completed 10 cycles between room
temperature and 100 $K$, and find no clear evidence that the diffraction
efficiency changes with temperature or with successive thermal cycle.
\end{abstract}

\section{Introduction}

Volume-Phase Holographic (VPH) gratings potentially have many advantages
over classical surface-relief gratings (Barden, Arns \& Colburn 1998;
Barden et al. 2000). They are already in operation in some existing
astronomical spectrographs (Kashiwagi et al. 2004) and their use is also
planned for a number of forthcoming instruments (Smith et al. 2004).
While the main applications of VPH gratings are currently for optical
spectrographs, they will also be useful for near-infrared (NIR)
spectrographs if the performance at low temperatures is satisfactory. In
particular, diffraction efficiency and its dependency on wavelength
should be confirmed. Contraction of dichromated gelatin with decreasing
temperature could cause variations in line density and the diffraction
efficiency profile (the thickness of the gelatin layer is one of the
parameters that define diffraction efficiency). Since thermal cycling
may cause some deterioration of a VPH grating and reduce its operational
life time, we also need to investigate whether the characteristics vary
with successive thermal cycles. Previously, we tested a VPH grating at
200 $K$ and confirmed that the performance is nearly independent of
temperature during 5 thermal cycles (Tamura et al. 2003, 2004). While
cooling to 200 $K$ can be sufficient for a spectrograph operating at
wavelengths out to $\sim$ 1.8 $\mu$m ($H$ band) such as Fiber Multi
Object Spectrograph (FMOS; e.g. Kimura et al. 2003), a much lower
temperature (e.g., 100 $K$) is required to extend the spectral coverage
of a spectrograph out to $\sim$ 2.4 $\mu$m ($K$ band).

\begin{figure}
\begin{center}
 \includegraphics[width=12cm,keepaspectratio]{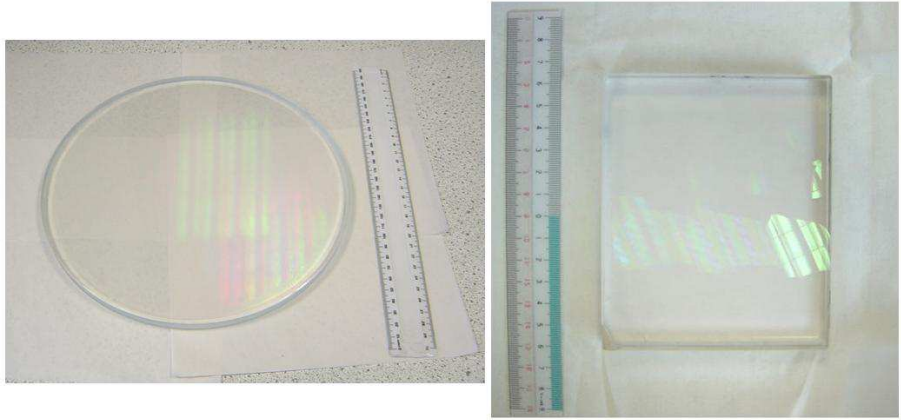}
 \caption{(Color online) Pictures of the VPH gratings investigated in
 this paper. The grating optimized for NIR is shown in the left
 panel. It has a diameter of 250 mm and a line density of 385 lines/mm.
 The grating for visible is shown in the right panel. It has a
 size of 100 mm $\times$ 120 mm and a line density of 300
 lines/mm.}
 \label{vphpics}
\end{center}
\end{figure}

In this paper, measurements of diffraction efficiency of VPH gratings at
100 $K$ and at room temperature are reported. Pictures of the gratings
investigated are shown in Fig. \ref{vphpics}, both of which were
manufactured by Ralcon Development Lab. The one in the left panel of
Fig. \ref{vphpics} has a line density of 385 lines/mm and the peak of
diffraction efficiency exists around 1.3 $\mu$m at the Bragg condition
when the incident angle of an input beam to the normal of the grating
surface is 15$^{\circ}$. The measurements of this grating are performed
at wavelengths from 0.9 $\mu$m to 1.6 $\mu$m. Since it is important to
see whether the performance at low temperatures is different from
grating to grating (Bianco et al. 2003; Blais-Ouellette et al. 2004;
Branche et al. 2004) we also investigate a different VPH grating, which
is shown in the right panel of Fig.  \ref{vphpics}. (This grating was
provided as a free sample for demonstration purposes only.)
The line density is 300 lines/mm and the peak efficiency is obtained
around 0.7 $\mu$m for an incident angle of 6$^{\circ}$. The measurements
of this grating are performed at wavelengths from 0.5 $\mu$m to 0.9
$\mu$m.

\section{The test facility and measurements}

%\begin{figure}[t]
%\begin{center}
% \includegraphics[height=10cm,angle=-90,keepaspectratio]{system0.ps}
% \caption{A picture of the test facility.}  \label{system0}
%\end{center}
%\end{figure}

\begin{figure}
\begin{center}
 \includegraphics[width=12cm,keepaspectratio]{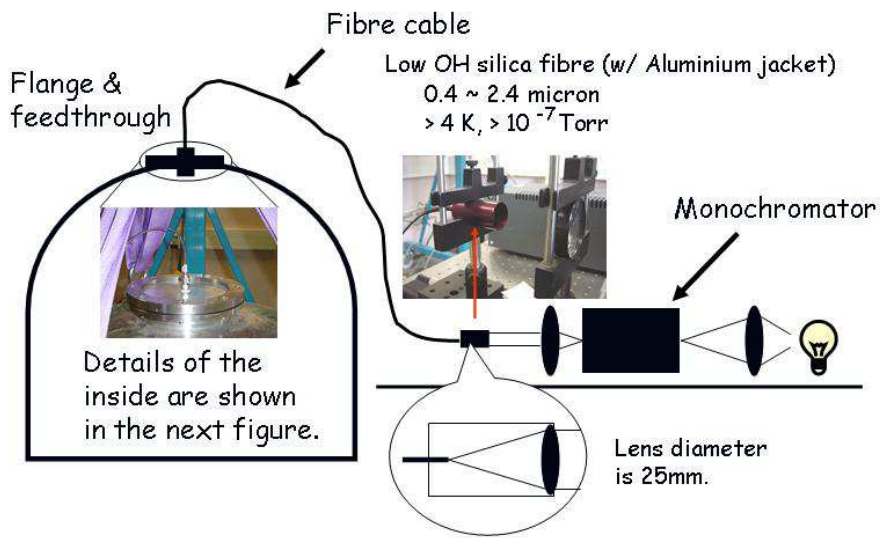}
 \caption{(Color online) Schematic view of the fore optics. The inside
 of the cryogenic chamber is described in Fig. \ref{system2}.}
 \label{system1}
\end{center}
\end{figure}

\begin{figure}
\begin{center}
 \includegraphics[width=12cm,keepaspectratio]{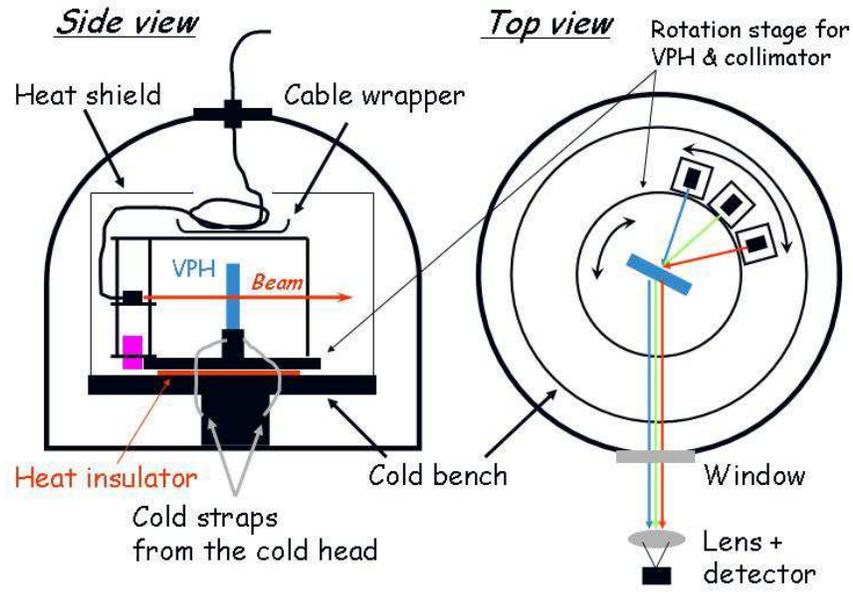}
 \caption{ (Color online) Schematic view of the inside of the cryogenic
 chamber. To the right, the mechanism to scan the incident beam angle to
 the VPH grating and wavelength is described.} \label{system2}
\end{center}
\end{figure}

In Fig. \ref{system1} and \ref{system2}, the overall configuration of
the optical components used for the measurements is indicated (detailed
information for the main components is given in Table \ref{comps}).
Light exiting from the monochromator is fed into the cryogenic chamber
through the fiber cable (Fig. \ref{system1}). Lenses with the same
specifications are attached to both ends of this fiber cable so that a
collimated beam received at one end exits from the other end in the
cryogenic chamber. This beam is directed towards the central axis of the
cold bench with parallel to the bench and illuminates the central part
of the VPH grating at a controlled incident angle. The spectral
band-width of this input beam is set by the slit width at the exit of
the monochromator. The slit width and the corresponding spectral
band-width are set to 0.5 mm and $\sim$ 0.01 $\mu$m, respectively,
throughout the measurements. The input beam diameter is $\sim 25$ mm
which is defined by the lens diameter of the fiber cable assembly. The
incident beam is then diffracted by the grating. The position of the
output collimator lens and the VPH grating are rotated independently
around the central axis of the cold bench so that the diffracted beam
for a certain combination of incident angle and wavelength goes through
the window towards the camera section consisting of lenses and a
detector (Fig.  \ref{system2}). The output fiber core is thus re-imaged
on the detector.

The basic measurement procedures are as follows: First, the brightness
of the lamp and the wavelength of light exiting from the monochromator
are fixed (the brightness of the lamp is kept constant by a stabilized
power supply), and the total intensity included in the image of the
fiber core is measured without a VPH grating. This measurement is
performed at all the sampling wavelengths and is also carried out both
at room temperature and cold temperature. The brightness of the lamp can
be changed when moving from one wavelength to another; a higher
brightness is used at the shortest and longest wavelengths because the
system throughput is lower. These data are used to normalize the
intensities measured when a VPH grating is inserted into the test set-up
and to calculate the diffraction efficiency. Since marginal differences
($\sim 2$\%) were found in the intensities measured without a VPH
grating between room temperature and cold temperature, we use the
intensities taken at room (cold) temperature to normalize those measured
with a VPH grating at room (cold) temperature, respectively.

\begin{figure}
\begin{center}
 \includegraphics[height=7cm,angle=-90,keepaspectratio]{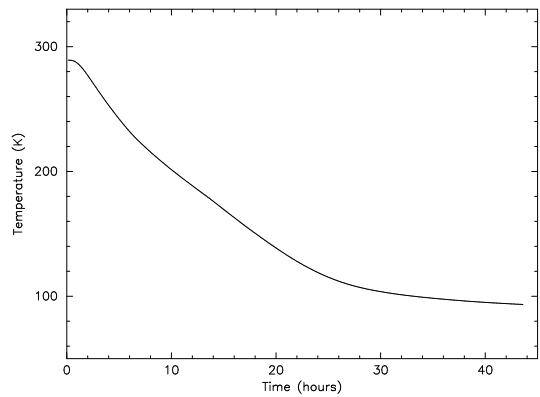}
 \caption{Temperature variation of the VPH grating during a typical
 cooling cycle.} \label{tempmon}
\end{center}
\end{figure}

Next, a VPH grating is inserted and the intensity of the first order
($m=+1$) diffracted light is measured for an incident angle of
10$^{\circ}$, 15$^{\circ}$, and 20$^{\circ}$ at all the sampling
wavelengths. Once all these measurements are performed, a cooling cycle
of the VPH grating is started. The temperature of the grating is
monitored with a calibrated silicon diode sensor on the grating surface,
close to the edge of the grating but unilluminated by the input beam.
For a good thermal contact between the grating and the sensor and
accurate measurements of the grating temperature, we put a thin layer of
grease to increase the contact area between the two surfaces. We also
use a device to keep pushing the sensor against the grating lightly
during a thermal cycle, which is thermally insulated from the metal
components of the test facility.
An example of the temperature variation is shown in Fig. \ref{tempmon}.
When the temperature of the grating becomes lower than 100 $K$, we start
the same sequence of measurements as above with running the compressor
and cold heads. There is no closed loop control of the grating
temperature as the rate of temperature variation is very low at 100
$K$. The temperature of the grating therefore stays approximately at
$\sim$ 90 $-$ 100 $K$ for the duration of the measurements. After the
measurements at $\sim$ 100 $K$, the compressor and cold heads are
switched off and the VPH grating is allowed to warm up passively. The
measurements are repeated when the temperature is back to the ambient
temperature. During these thermal cycles, the cryogenic chamber is kept
evacuated to $\sim$ 10$^{-7}$ Torr.

\section{Results and discussions}

\begin{figure}
\begin{center}
 \includegraphics[height=13.2cm,angle=-90,keepaspectratio]{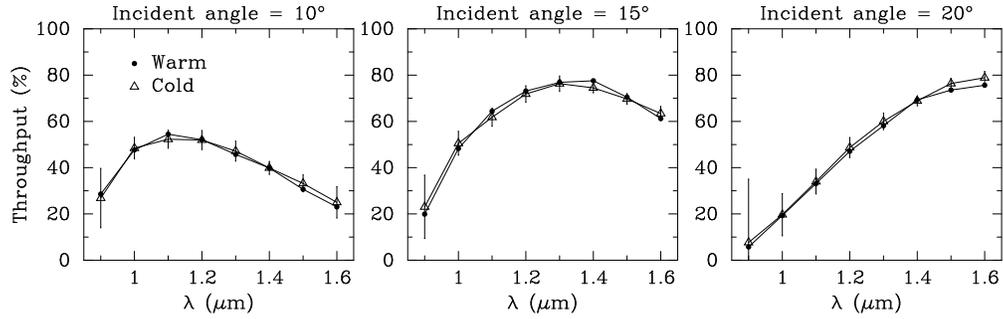}
 \caption{Diffraction efficiency of the NIR VPH grating measured for an
 incident angle of 10$^{\circ}$ (left), 15$^{\circ}$ (center) and
 20$^{\circ}$ (right).  These plots show the measurements of the first
 ($m = +1$) order diffraction.} \label{effnir}
\end{center}
\end{figure}

First, we show results from the VPH grating for NIR wavelengths. In
Fig. \ref{effnir}, measured efficiency of the first order ($m=+1$)
diffraction is plotted against wavelength for an incident angle of
10$^{\circ}$, 15$^{\circ}$, and 20$^{\circ}$ in the left, middle, and
right panel, respectively. The circles indicate the data points at room
temperature, and the triangles are those at 100 $K$.
%The upper plots show the measurements of the first ($m=+1$) order
%diffraction, and the lower plots indicate those of the zero order light.
The error bars indicate estimated random errors in the measurements
expected due to pixel-to-pixel variation of background intensities in
the images of the fiber core and subsequent uncertainty of background
subtraction. Note that the large error at $0.9$ $\mu$m is due to the
lower sensitivity of the detector at the edge of the spectral coverage.
For clearer presentation, these error bars are attached only to the data
points for a cold test, but those for a warm test are similar. These
results suggest that a VPH grating can withstand cryogenic temperatures
in vacuum
%due to, e.g., crystallization and outgas
and that its performance at 100 $K$ is similar to that at room
temperature.

\begin{figure}
\begin{center}
 \includegraphics[height=6cm,angle=-90,keepaspectratio]{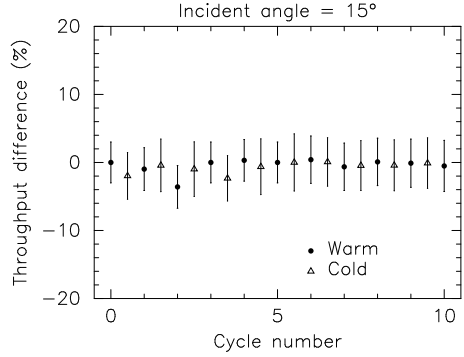}
 \caption{The differences in diffraction efficiency of the NIR VPH
 grating at one test from the first warm test are averaged over the
 wavelength range investigated and plotted against cycle number. Open
 triangles and solid dots represent the data at 100 $K$ and those at
 room temperature, respectively. The error bars indicate a combination
 of the standard deviation of a distribution of the differences around
 the average value and the typical uncertainty in the measurement of
 diffraction efficiency.}  \label{effvarnir}
\end{center}
\end{figure}

In order to see whether the performance of this VPH grating deteriorated
with successive thermal cycling, the differences in diffraction
efficiency for an incident angle of 15$^{\circ}$ between the first warm
test and subsequent tests are averaged over the wavelength range
investigated and
%this averaged difference is 
plotted against cycle number in Fig. \ref{effvarnir}. Circles and
triangles indicate the measurements at room temperature and those at 100
$K$, respectively. The error bar represents a combination of the
standard deviation of the differences around the average value and the
typical uncertainty ($\sim$ 3 \%) in the measurement of diffraction
efficiency. Similar results are obtained from the data for the other
incident angles (10$^{\circ}$ and 20$^{\circ}$). This result suggests
that no significant deterioration of a VPH grating is caused by thermal
cycling.

\begin{figure}
\begin{center}
 \includegraphics[height=13.2cm,angle=-90,keepaspectratio]{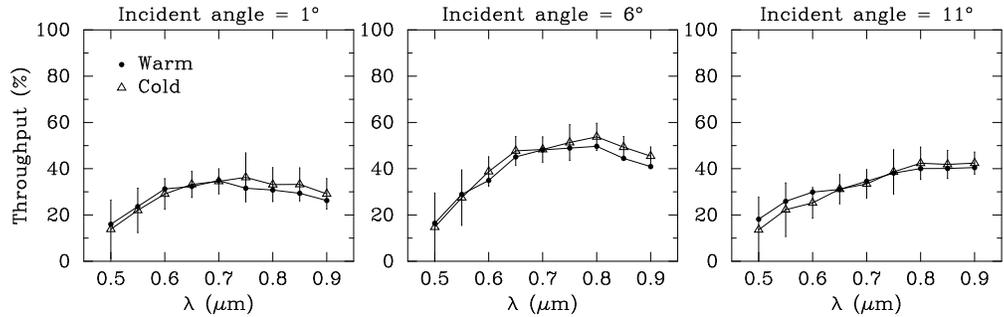}
 \caption{Same as Fig. \ref{effnir}, but for the VPH grating for visible
 wavelengths. The measured diffraction efficiency for incident angles of
 1$^{\circ}$, 6$^{\circ}$, and 11$^{\circ}$ is plotted against
 wavelength.}  \label{effvis}
\end{center}
\end{figure}

\begin{figure}
\begin{center}
 \includegraphics[height=6cm,angle=-90,keepaspectratio]{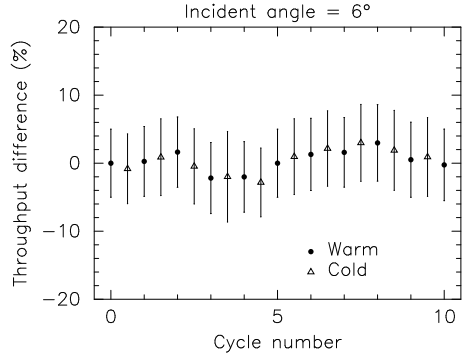}
 \caption{Same as Fig. \ref{effvarnir}, but for the VPH grating for
 visible wavelengths.} \label{effvarvis}
\end{center}
\end{figure}

Next, we present results from the VPH grating for visible
wavelengths. The measurement procedure is the same as that in the NIR
except that a CCD camera is used and the grating is accommodated with a
different mount on the cold bench. In Fig. \ref{effvis}, the
measurements of diffraction efficiency at 100 $K$ are compared with
those at room temperature for an incident angle of 1$^{\circ}$,
6$^{\circ}$, and 11$^{\circ}$ in the left, middle, and right panel,
respectively.
%Note that the measurements are slightly more uncertain than those in the
%NIR experiments mainly because of the low throughput of the grating and 
%high background noise of the CCD camera.
In Fig. \ref{effvarvis}, the average difference from the first warm test
is plotted against cycle number for an incident angle of 6$^{\circ}$.
These results again suggest that the performance does not largely depend
on temperature or the number of thermal cycles. We note that the
throughput is significantly lower than the NIR grating. The reason for
this is unknown, but this grating was for demonstration purposes only
and hence the low throughput may be due to, e.g., severe internal
absorption and/or imperfect fringes.

\begin{figure}
\begin{center}
 \includegraphics[height=6cm,angle=-90,keepaspectratio]{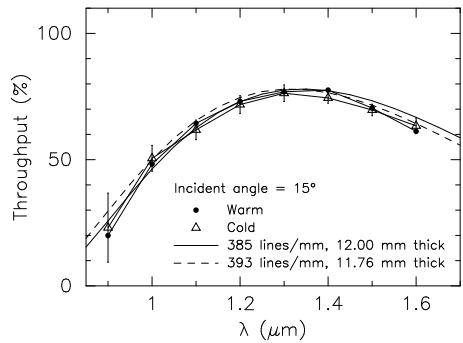}
 \caption{Diffraction efficiency of the NIR VPH grating for an incident
 angle of 15$^{\circ}$ is plotted as a function of wavelength. Black
 solid line indicates diffraction efficiency calculated with coupled
 wave analysis for a line density of 385 lines/mm and a gelatin
 thickness of 12.00 $\mu$m. Dashed line indicates diffraction efficiency
 computed for a line density of 393 lines/mm and a gelatin thickness of
 11.76 $\mu$m, representing that expected at 100 $K$. In both
 calculations, a fringe amplitude of 0.055 is assumed. These model
 predictions are compared with the actual measurements shown in the
 middle panel of Fig. \ref{effnir}, which are plotted with with solid
 circles and open triangles.} \label{change}
\end{center}
\end{figure}

This robustness of diffraction efficiency to temperature variation is
expected for these VPH gratings in theory, provided that
%the property of a VPH grating does not dramatically change at low
%temperature due to, e.g., crystallization of gelatin, and that
the linear thermal expansion coefficient of gelatin at 100 $-$ 200 $K$
is similar to that at 200 $-$ 300 $K$, i.e., in the range of $10^{-4} -
10^{-5}$ $K^{-1}$. Given a linear thermal expansion coefficient of
10$^{-4}$ $K^{-1}$, a gelatin thickness would be 2 \% smaller at 100 $K$
compared to that at 300 $K$. Considering the VPH grating for NIR
wavelengths and assuming a 12 $\mu$m thickness at room temperature,
which gives a good fit of the predicted throughput curve to the
measurements for this NIR VPH grating (Tamura et al. 2003), the
thickness at 100 $K$ would be 11.76 $\mu$m. Also, the line density would
be increased by the same fraction. Since the line density of this
grating is 385 lines/mm in the specification, it would be 393 lines/mm
at 100 $K$. In Fig. \ref{change}, the diffraction efficiency predicted
with coupled wave analysis (Kogelnik 1969) is plotted against wavelength
for the two sets of line density and gelatin thickness; one is 385
lines/mm and 12 $\mu$m, and the other is 393 lines/mm and 11.76
$\mu$m. The fringe amplitude in refractive index is assumed to be 0.055
in both calculations. An incident angle of 15$^{\circ}$ is also assumed,
but the difference between the two calculations is similarly small for
incident angles of $10^{\circ}$ and $20^{\circ}$. Note that the
predicted diffraction efficiencies are scaled by a factor of 1.2 to fit
them to the measurements, indicating that the throughput is $\sim 20$ \%
lower than the theoretical prediction. About half of this discrepancy
can be explained by energy loss due to reflections at interfaces between
glass substrate and ambient space. The other half has not been
identified but it perhaps includes, e.g., internal absorption
(identifying the source of this energy loss is beyond the scope of this
paper). These calculations suggest that the expected change in
throughput is as small as confirmed by the measurements.

\section{Summary \& conclusion}

In this paper, we present results from cryogenic tests of VPH gratings
at $\sim$ 100 $K$. The aims of these tests are to see whether the
diffraction efficiency at a low temperature as a function of wavelength
is significantly different from that at room temperature and to see
whether the grating can withstand a number of thermal cycles. Having
exposed VPH gratings to 10 cycles between room temperature and 100 $K$,
we find that diffraction efficiency measured at 100 $K$ agrees with that
at room temperature within the errors. We also find no clear evidence
that the performance changes with the successive thermal cycles. These
results were found for both of the two different VPH gratings
investigated here, which may imply that VPH gratings can withstand such
cryogenic temperatures in general. Ideally, an investigation of more
gratings, in particular from different manufacturers and with different
substrate materials, should be carried out to confirm this point.

It needs to be emphasized that we have only confirmed the performance of
a low dispersion VPH grating at cryogenic temperature. It would be
useful to repeat the same experiments for high dispersion VPH
gratings. Since the band-width of a throughput curve is narrower for
high dispersion gratings, some changes of physical properties (e.g.,
gelatin thickness) due to temperature variations are expected to be
revealed much more clearly in the form of a shift of the throughput peak
and/or a global decrease of the throughput. In this case, one would have
to predict the change of characteristics due to temperature variations
and take them into account in the design and fabrication of a VPH
grating so that it could work with the optimal performance at the
operating temperature.

\section*{Acknowledgements}
We thank colleagues in Durham for their assistance with this work,
particularly J\"{u}rgen Schmoll, Daniel Gedge, and the members of the
mechanical workshop. We are also grateful to Ian Parry for letting us
use his VPH grating for visible wavelengths. This work was funded by
PPARC Rolling Grant PPA/G/O/2003/00022.

\setlength{\tabcolsep}{1.5mm}
\begin{table}
\begin{center}

%\caption{The main components used for the measurements.}
Table 1: The main components used for the measurements.

\vspace{2mm}

\begin{tabular}{llll} \hline\hline
                       & Manufacturer & Product ID & \multicolumn{1}{c}{Comments} \\ \hline
Light source           & Comar                 & 12 LU 100        & Tungsten-halogen lamp \\
Monochromator          & Oriel Instruments     & Cornerstone 130, & 600 lines/mm grating, \\
                       &                       & Model 74000      & Blaze at 1 $\mu$m \\
Fiber cable            & Fiberguide Industries & Custom assembly  & $\lambda\lambda$: 0.4 $-$ 2.4 $\mu$m \\
                       &                       &                  & $> 4~K$, $> 10^{-7}$ Torr \\
\multicolumn{1}{l}{\hspace{4mm}- Fiber} &      & AFS200/240A      & Aluminium-jacketed \\
                       &                       &                  & low OH silica fiber \\
\multicolumn{1}{l}{\hspace{4mm}- Collimators} && Custom assembly  & Plano convex lenses \\
                       &                       &                  & (BK7, $\phi = 25$mm) \\
                       &                       &                  & at both ends of the cable \\
Near-infrared detector & Indigo Systems        & Alpha$-$NIR      & 320 $\times$ 256 InGaAs array \\
CCD camera             & Starlight Xpress      & MX516            & 500 $\times$ 290 pixels \\ \hline\hline
\end{tabular}
\label{comps}
\end{center}
\end{table}

\clearpage

\end{document}